\documentclass{article}
\usepackage{spconf,amsmath,mathtools,graphicx,cite,algorithm2e,acronym}
\usepackage{subfigure}

\graphicspath{{figure/}}
\title{Graph-Based Recommendation System}

\name{Kaige Yang, Laura Toni}
\address{ Dept. of   Electronic and Electrical Engineering\\
University College London, London, UK \\
	\fontsize{9}{9}\selectfont\ttfamily\upshape \{kaige.yang.11,\,l.toni\}@ucl.ac.uk}

\begin{document}
%\ninept
%
\maketitle
\begin{abstract}
In this work, we study recommendation systems  modelled as contextual multi-armed bandit (MAB) problems. We propose   a graph-based recommendation system that learns and exploits the   geometry of the  user space to create meaningful clusters in the user domain. This   reduces the dimensionality of the recommendation problem while preserving the  accuracy of   MAB.  We then study the effect of graph sparsity and   clusters size  on the MAB performance and provide exhaustive simulation results   both  in synthetic and in real-case datasets. Simulation results show improvements with respect to state-of-the-art MAB algorithms.

\end{abstract}

\begin{keywords}
Recommendation system, contextual multi-armed bandit, community detection.
\end{keywords}

\section{Introduction}
\label{sec:intro}
Recommending products to users have been an essential function of commercial websites as Amazon and Netflix, etc. \cite{amazon}. The aim of a recommendation agent is to propose to a user the product (or item) that will generate a positive reaction: a product purchase in Amazon, a link click in ads website, etc. This user response increases the agent payoff, which ultimately needs to be   maximized. 
The effectiveness of a recommendation system depends on the knowledge of users' preference: the deeper the knowledge, the more the tailored is the  recommended item. The challenge is that these preferences usually are not known a priori and need to be  built online by trial and error for each user.  
This learning process can be formalised by multi-armed bandit (MAB) framework\cite{Robbins1952, chen2013personalized,liu2010personalized,kim2009personalized}. 

The performance of MAB learning strategies 
scales with the ambient  dimension, either linearly or as a square root \cite{bubeck2012regret}, which makes the problem intractable in scenarios with infinitely large strategy sets, as in recommendation systems. 

To overcome the dimensionality limitation, clustering techniques have been proposed to properly quantize the context space (e.g., the user space)~\cite{slivkins2014contextual}. 
Users preference relationships can   be   encoded in a graph, where adjacent nodes represents users with similar preferences~\cite{Nguyen:2014,rezaeimehr2018tcars}. This graph can be known a-priori  or it can be   inferred   based on the past users' feedbacks (past payoffs). We are interested in this latter case and in recent works in which the  geometrical and irregular structure  of the context have been considered~ \cite{rezaeimehr2018tcars,Gentile2014,li2015data, cesa2013gang,LiGKZ16,li2016collaborative,korda2016distributed,Caron:A12}.  In \cite{Gentile2014},  authors proposed CLUB, an online clustering strategy where $m$ clusters are optimized for $n$ ($>>m$) users, where only one recommendation per cluster is optimized.  In    \cite{li2016collaborative}, a similar  idea has been implemented on both user and item side to propose COFIBA.  In both CLUB and COFIBA, an iterative graph learning process is considered with a fully connected graph as starting point. At each recommendation opportunity, edges are deleted if connecting users with different enough payoff. Any connected component will then form a cluster.  The leads to a very simple and yet effective MAB problem, but with the limitation of a $i)$ limited clustering strategy, $ii)$ no  possibility of recovering from inaccurate payoff estimate (edges can only be deleted from the known graph and they cannot be added in case of edges wrongly deleted in the past), and $iii)$ a number of clusters rapidly increasing with time, which we show not to be   the best trend for MABs.     
In contrast,  DYnUCB \cite{nguyen2014dynamic} groups users via K-means assigning users dynamically into clusters. 
However, it requires an input as a pre-defined number of clusters. While theoretically  $K$ can be optimized with iterative solutions (e.g., elbow method), in practice, an appropriated cluster number is typically unknown, hard to guess, and dynamic over time (as users might appear or disappear). 

To overcome these limitations, in this paper we propose   SCLUB-CD,  a novel graph-based MAB problem that learns and exploits the geometry of the user domain in an online fashion. Specifically, at each recommendation opportunity, a user graph is constructed based on estimated user preference. Then, the graph is  divided into clusters based on the community detection algorithm~\cite{blondel2008fast}. 
Our main contributions are:
\begin{itemize}
    \item to adopt graph clustering into MABs to propose a dynamic graph estimation \emph{and}  clustering. 
    \item to show that MABs are more efficient when the number of cluster  remains limited over time. Therefore, our proposed recommendation system keeps the number of clusters limited over time,  opposite behaviour  with respect to CLUB.  
    \item to  test the proposed algorithm in both synthetic and realist dataset, showing  improved performance with respect to LinUCB and CLUB state-of-the-art MAB problems.
\end{itemize}
%%%%%%%%%%%%%%%%%%%%%%%%%%%%%%%%%%%%%%
\section{MAB for Recommendations}
\label{sec:setting}
We now describe the basics on recommendation systems and how these problems, when tackled as MAB algorithms, can benefit from context clustering. 

Let each user $i\in \mathcal{I}$ be identified by its preferences  $\boldsymbol{u}_i \in \mathbf{R}^l$
, and let the  product  $k\in \mathcal{K}$ be identified  by its  own feature vector   $\boldsymbol{x}_k\in \mathbf{R}^d$ (identifying information such as size, colour and price of the product), with   $l,d$ being the dimension of the user and product vector, respectively, and    $|\mathcal{I}|=N$. While $ \boldsymbol{x}_k$ is known to the agent, the user preference vectors  $\boldsymbol{u}_i$ need to be learned on the fly. To this effect, the agent makes sequential recommendation and observes the outcome (appreciation of the recommended product). Formally, at each recommendation opportunity $t=1,2,...,T$,   the agent receives a user index $i_t \in \mathcal{I}$   to serve content to, with $i_t$  selected uniformly at random from  $\mathcal{I}$. It also receives the set of possible products to recommend  $\mathcal{C}_t \subset \mathcal{K}$, with $|\mathcal{C}_t|=C$. The agent then recommends one product out of the available ones to user $i_t$ and observe the user's feedback in the form of   instantaneous payoff $a_t$. The payoff is assumed to be a linear function of the product features $\boldsymbol{x}_k$ and user preference vector $\boldsymbol{u}_{i_t}$ with a noise term $\epsilon \sim \mathcal{N}(0,\sigma^2_{\epsilon})$.~\cite{liu2010personalized}. Namely, 
\begin{equation}
\label{eq:payoff}
a_t={\boldsymbol{u}_{i_t}}^T\boldsymbol{x}_k+\boldsymbol{\epsilon}_{i_t},    \ \ \ \ 
 k \in \mathcal{C}_t   
\end{equation}
with $a_t \in [0,1]$, with $1$ being the highest appreciation and $0$ the lowest, and $\boldsymbol{\epsilon}_{i_t}(\boldsymbol{x})$ being a random Gaussian noise $\epsilon\sim\mathcal{N}(0,\sigma^2_{\epsilon})$.

Note that  ${\boldsymbol{u}_{i_t}}^T\boldsymbol{x}$ is the expected payoff received from user ${i_t}$ for  $\boldsymbol{x}$, while $a_t$ is the instantaneous one.

Let us   assume that users are clustered in $M$ non overlapping clusters based on their preferences,  with  $V_j$, $j=1,2,..,M$ being the $j$th cluster, and   $M$ be unknown a priori.  Due to the linear payoff scenario, users in the same clusters will experience similar payoff functions. It follows that rather than having a preference vector per user, the agent can identify and learn a preference vector per cluster. Therefore, the agent needs to learn only $M$ preferences vector rather than $N$, with $N \gg M$. This comes at the price of   an approximation in the estimation of the linear payoff, and therefore a suboptimality in the recommendation.  More formally, each user cluster $V_j$ has the preference vector $\boldsymbol{u}_j^{\text{c}}$ representing the common parameter vector shared by users within the cluster, leading to an \emph{estimated} mean  payoff given by  
${\boldsymbol{u}^{\text{c}}_{j(i_t)}}^T\boldsymbol{x}_k+\boldsymbol{\epsilon}_{j(i_t)},$ with $ k \in \mathcal{C}_t 
$ and  $j(i_t)$ being the cluster index whose  user $i$ belongs to. Note that the actual payoff (per user) is given by \eqref{eq:payoff}, while the agent will estimate the above one per cluster. From here the suboptimality of the clustering-based recommendations. 

The agent aims to minimise the cumulative regret  $R_T$ over the time horizon $T$ defined as $R_T=\sum^T_{t=1}r_t,$ with  $r_t$ being the regret at time $t$, defined as the difference between the payoff incurred by the algorithm and the optimal payoff. Formally, 

\begin{align}
    r_t&=\max_{k\in\mathcal{C}_t}\{{\boldsymbol{u}_{i_t}^T\boldsymbol{x}}_{k}\}-{\boldsymbol{u}_{i_t}}^T\boldsymbol{x}_t
\end{align}
where, $r_t$ is the regret at time $t$.   Following the MAB theory, the cumulative regret is minimized if products are selected as follows
\begin{equation}
\label{eq:UCB}
    k_{t}=\underset{k\in\mathcal{C}_t}{\arg\max}({\boldsymbol{u}}^{\text{c}}_{j(i_{t})}\boldsymbol{x}_{k})+CB_{j(i_t)}(\boldsymbol{x}_{k})
\end{equation}
The quantity $CB_{j(i_t)}$ is   the upper confidence bound  of each arm with respect to cluster ${j(i_t)}$. Basically, a product is selected if the expected payoff is high and it is low the uncertainty on this estimated payoff. 
  
%%%%%%%%%%%%%%%%%%%%%%%%%%%%%%%%%%%%%%%%%%%%%
\section{Graph-Based MAB}
\label{sec:algorithm}
We now describe the proposed SCLUB-CD algorithm, depicted   in Algorithm 1.  

At the recommendation opportunity $t$, the agent estimates an
unweighted and undirected graph $G_t=(\mathcal{V},\mathcal{E}_t,  {W}_t)$,   with $\mathcal{V}$ being the vertex set representing the $N$ users\footnote{Without loss of generality, we assume the number of active users constant over time.}  with  $|\mathcal{V}|=N$, $\mathcal{E}_t$ and $W_t$  the edge sets  and the $N \times N$ adjacency matrix estimated at $t$, respectively. 

The  graph $G_t$ is obtained by following a 3 steps iterative method:

{\bf STEP 1.}
 First  an undirected and \emph{weighted} graph $\tilde{G_t} =(\mathcal{V},\mathcal{E}_t,  \tilde{W}_t)$ is estimated. Given the current knowledge of the system,   users preferences are estimated minimizing  the linear  least-square estimate of $\boldsymbol{u}$ as 
\begin{equation}
    \hat{\boldsymbol{u}}_{i,t}=M_{i,t}^{-1}\boldsymbol{b}_{i,t},  \ \ i=1,2,...,N
\end{equation}
Then, the graph weights $\tilde{\mathbf{w}}_{t,i,j}$ in $\tilde{G_t}$  are evaluated as  the Gaussian $\textit{RBF-distance}$ between $\hat{\boldsymbol{u}}_{i,t}$ and $\hat{\boldsymbol{u}}_{j,t}$.

{\bf STEP 2.}
Then, the graph is converted in a sparse and unweighted graph. Sparsity is motivated by the need to tune the dimensionality of the user space, while the binary weights are introduced mainly to increase the robustness of the algorithm to graph estimation errors. Both aspects will be discussed in the results section. 

We introduce the hyperparameter $n$, such that the top $n$, $n\ll N$, largest weights in $w_{t,i,j}, j=1,2,...,N$ are encoded as 1, the remaining are 0. The result graph is $G_t$. Note that we do not impose sparsity by  setting to zero all weights below a given threshold value (more common approach) but we introduce $n$ instead. This is to better control the number of clusters, that needs to remain low for an efficient learning.

{\bf STEP 3.}
Once the graph   $G_t$ is estimated, the clusters $\hat{V}_{1,t}$,$\hat{V}_{2,t}$,...,$\hat{V}_{M,t}$ are derived via community detection     applying the Louvain Method \cite{blondel2008fast}. Preferences per cluster are then estimated (as shown in Algorithm 1) and   the algorithm   selects the product $k_{t}$ at opportunity $t$ following the UCB method, i.e., following the minimization in \eqref{eq:UCB}. Once the product indexed by $k_{t}$ with feature $\boldsymbol{x}_{k_t}$ is recommended, $\boldsymbol{x}_{k_t}$ is used to update $\hat{\boldsymbol{u}}_{i,t}$ along with the received payoff $a_t$ via a standard linear least-square approximation of $\boldsymbol{u}_{i_t}$ (as shown in Algorithm 1) and a new loop starts.

\begin{algorithm}[t]
\caption{SCLUB-CD Algorithm}
\scriptsize
\textbf{Initial}:
$b_{i,0}=\mathbf{0} \in \mathbf{R}^d $,
$M_{i,0}=\boldsymbol{I} \in \mathbf{R}^{d \times d}$,
$\hat{\boldsymbol{u}}_{i,0}=\mathbf{0} \in \mathbf{R}^d$, $i \in [1,N]$;\\
\textbf{Input}: Edge deletion parameter $n\in (0,N]$;\\
\For{$t=1,2,...,T$} 
{
    \textbf{Set} $\hat{\boldsymbol{u}}_{i,t-1}=M_{i,t-1}^{-1}\boldsymbol{b}_{i,t-1}, i=1,2,...,N$;\\
    \textbf{Find} $G_{t-1, weighted}$: 
    Calculate pairwise \textit{Gaussian RBF-distance} $\tilde{\mathbf{w}}_{t-1,i,j}$ between $\hat{u}_{t-1,i}$ and  $\hat{u}_{t-1,j}$, $i,j=1,2,...,N$;\\
    \textbf{Find} $G_{t-1}$:
    Set the largest $n$ values of 
    $\tilde{\mathbf{w}}_{t-1,i,j}, j=1,2,..,N$ as 1;
    Set the remaining to 0;\\
    \textbf{Find} $\hat{V}_{\hat{j}_t,t-1}$:
    Apply Louvain Method on $G_t$;\\
\textbf{Set}
    \begin{equation*}
        \bar{M}_{\hat{j}_{t,t-1}}=\boldsymbol{I}+\sum_{i\in\hat{V}_{\hat{j}_t,t-1}}(M_{i,t-1}-\boldsymbol{I})  
    \end{equation*}
    \begin{equation*}
        \bar{\boldsymbol{b}}_{\hat{j}_t,t-1}=\sum_{i\in\hat{V}_{\hat{j}_t,t-1}}\boldsymbol{b}_{i,t-1},
        \bar{\boldsymbol{u}}_{\hat{j}_t,t-1}=\bar{M}_{\hat{j}_t,t-1}^{-1}\bar{\boldsymbol{b}}_{\hat{j}_t,t-1}
    \end{equation*}
\textbf{Find}
    \begin{equation*}
        k_t=\underset{k\in1,...,C}{\arg\max}(\bar{\boldsymbol{u}}_{j(i_{t-1})}\boldsymbol{x}_{t,k})+CB_{\hat{j}_{t-1}}(\boldsymbol{x}_{t,k})
    \end{equation*}
    \begin{equation*}
        CB_{\hat{j}_{t-1}}(\boldsymbol{x})=\alpha\sqrt{\boldsymbol{x}^T\bar{M}_{\hat{j}_t,t-1}\boldsymbol{x}\log{(t+1)}}
    \end{equation*}
\textbf{Receive} $a_t \in [0,1]$.\\
\textbf{Update}:
    \begin{equation*}
        M_{i_t,t}=M_{i_t,t-1}+\boldsymbol{x}_t\boldsymbol{x}_t^T,
      \boldsymbol{b_{i_t,t}}=\boldsymbol{b_{i_t,t-1}}+a_t\boldsymbol{x}_t
    \end{equation*}
    \begin{equation*}
        M_{i,t}=M_{i,t-1}, \ \boldsymbol{b_{i,t}}=\boldsymbol{b_{i,t-1}} , i \neq i_t
    \end{equation*}
}
\end{algorithm}

\section{Simulation Results}
\label{sec:Experiments}
\subsection{Simulations Setup}
We carried out results both in a synthetic and a realistic dataset. The synthetic case allows us to simulate a scenario in which we can actually control the similarity among users, while the realistic dataset has been implemented to validate our algorithm  in real recommendation problems. 
In the synthetic case, $N=100$ users are clustered in $M=5$ clusters and  $|\mathcal{K}|=1000$ products are considered. Out of these $1000$ products, at each trial $t$,   a smaller   pool $C$ with size $C=25$ is chosen uniformly at  random  from $\mathcal{K}$ as candidate for the recommendation\footnote{This is a  common assumption in recommendation systems, therefore we apply it in both  synthetic and realistic scenarios.}.  We set the dimension of both the users and product features vectors to  $l=d=25$.  To control the similarity among users within  a cluster, intra-cluster noise $\sigma_{\text{c}}$ is introduced.   For each user $i$ belonging to  $V_j$, $\boldsymbol{u}_i$ is created by perturbing the $\boldsymbol{u}_{j(i)}^{\text{c}}$ with a white noise term drawn uniformly at random across from a zero mean normal distrbution with variance   $\sigma_{\text{c}}^2$. The lower $\sigma_{\text{c}}^2$ the more compact the clusters.  In the following simulation results, we consider both  $\sigma_{\text{c}}$ and  $\sigma_{\epsilon}$ to be in the range $[0.25, 0.5]$, where we recall that $\sigma_{\epsilon}$ is the standard deviation of the payoff.

For the real-world datasets, we consider \textbf{LastFM}, containing tags of artists and  record listened by users,  and \textbf{Delicious}, including URLs bookmarked and tags provided by users, \cite{cantador2011second}. While \textbf{LastFM}  represents a scenario named ``few-hits" where users' preference are coherent (therefore it is reasonable to assume that users can be clustered),  \textbf{Delicious}  represents a  ``many-hits" scenario in which users' preferences are diverse (therefore the clustering is a strong approximation). Simulation results are  averaged over $10$ runs and provided in the following.

\textbf{LastFM} and \textbf{Delicious} were processed following the same procedure in \cite{cesa2013gang} which we detail here. First, tags contains in datasets were breakdown into single words by removing underscore, hyphens and apexes. Second, tags that appear less than 10 times were removed. Third, all tags related to each specific item was formed as a TF-IDF vector to represent the item feature. To reduce the dimension, PCA was applied to TF-IDF vectors and only the top 25 principle components were retained.

The proposed SCLUB-CD\footnote{Available at https://github.com/LASP-UCL/KaigeYang/tree/graph-based-recommendation-system.} is compared with respect to the state-of-the-art algorithms, namely  LinUCB \cite{li2010contextual}, CLUB\cite{Gentile2014}. In the synthetic dataset,  we provide simulation results also for the  SCLUB-CD in the case in which user clusters is known, but user preference is unknown.  We label this method  SCLUB-CD-Correct and it represents a lower bound in terms of cumulative regret. To study the effect of the weigted and sparse graph, we provide results also for  two other baseline methods (modified version of the proposed SCLUB-CD):  SCLUB-CD-Weight performs the clustering based on $\hat{G}_t$, while SCLUB-CD-Weight-Sparse keeps the top $n$ largest Gaussian \textit{RBF-distance}  edges but preserving their weights.

%%%%%%%%%%
\subsection{Results}

\begin{figure} 
\centering
\subfigure[$\sigma_{\text{c}}=0.25$, and  $\sigma_{\epsilon}=0.25$.]{ \includegraphics[width=.47\linewidth,draft=false]{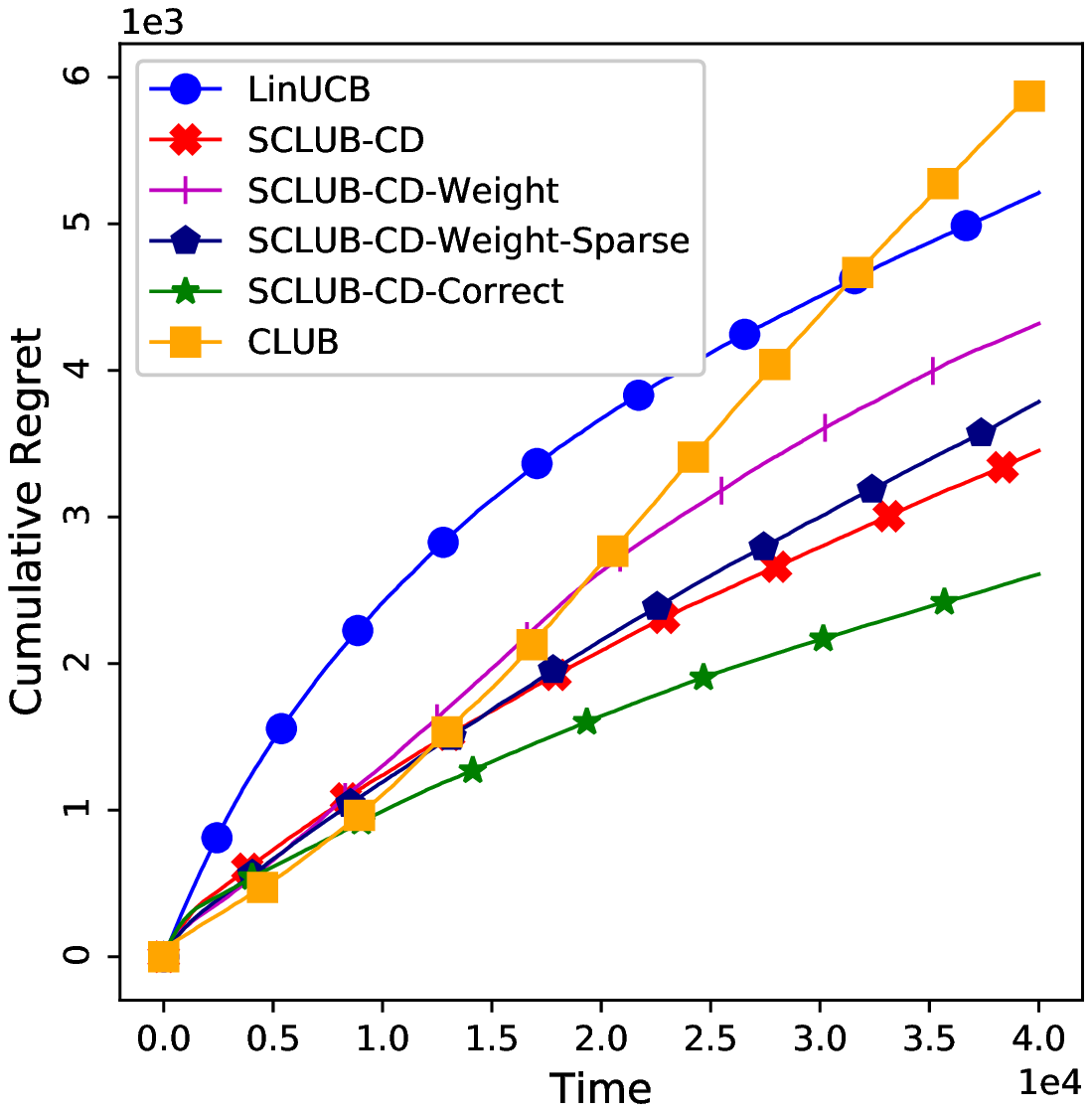}
\label{fig:data_performance_25_25}}
\hfill
\subfigure[$\sigma_{\text{c}}=0.5$, and  $\sigma_{\epsilon}=0.25$.]{ \includegraphics[width=.47\linewidth,draft=false]{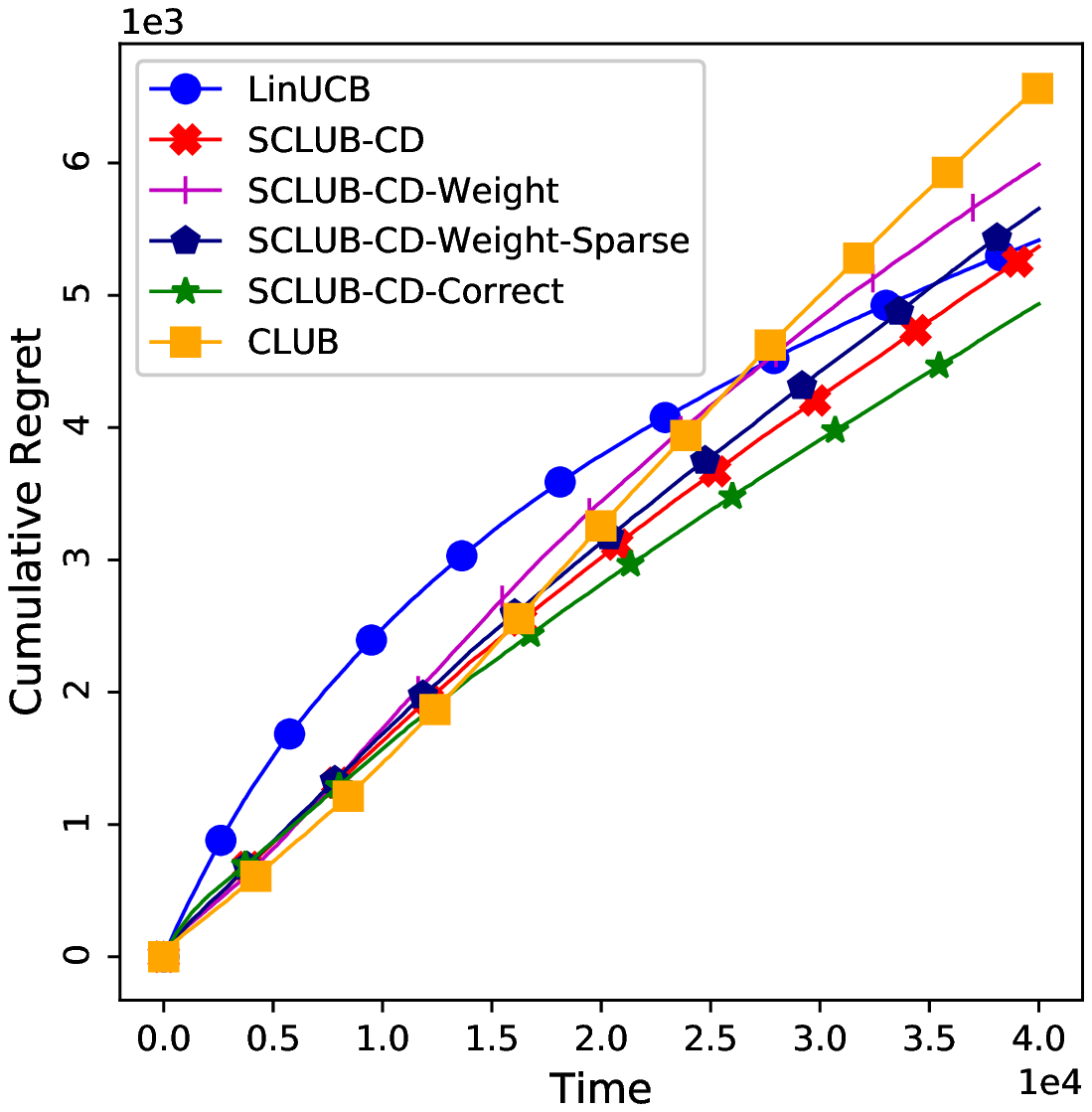}
\label{fig:data_performance_25_50}}
\subfigure[$\sigma_{\text{c}}=0.25$, and  $\sigma_{\epsilon}=0.5$.]{ \includegraphics[width=.47\linewidth,draft=false]{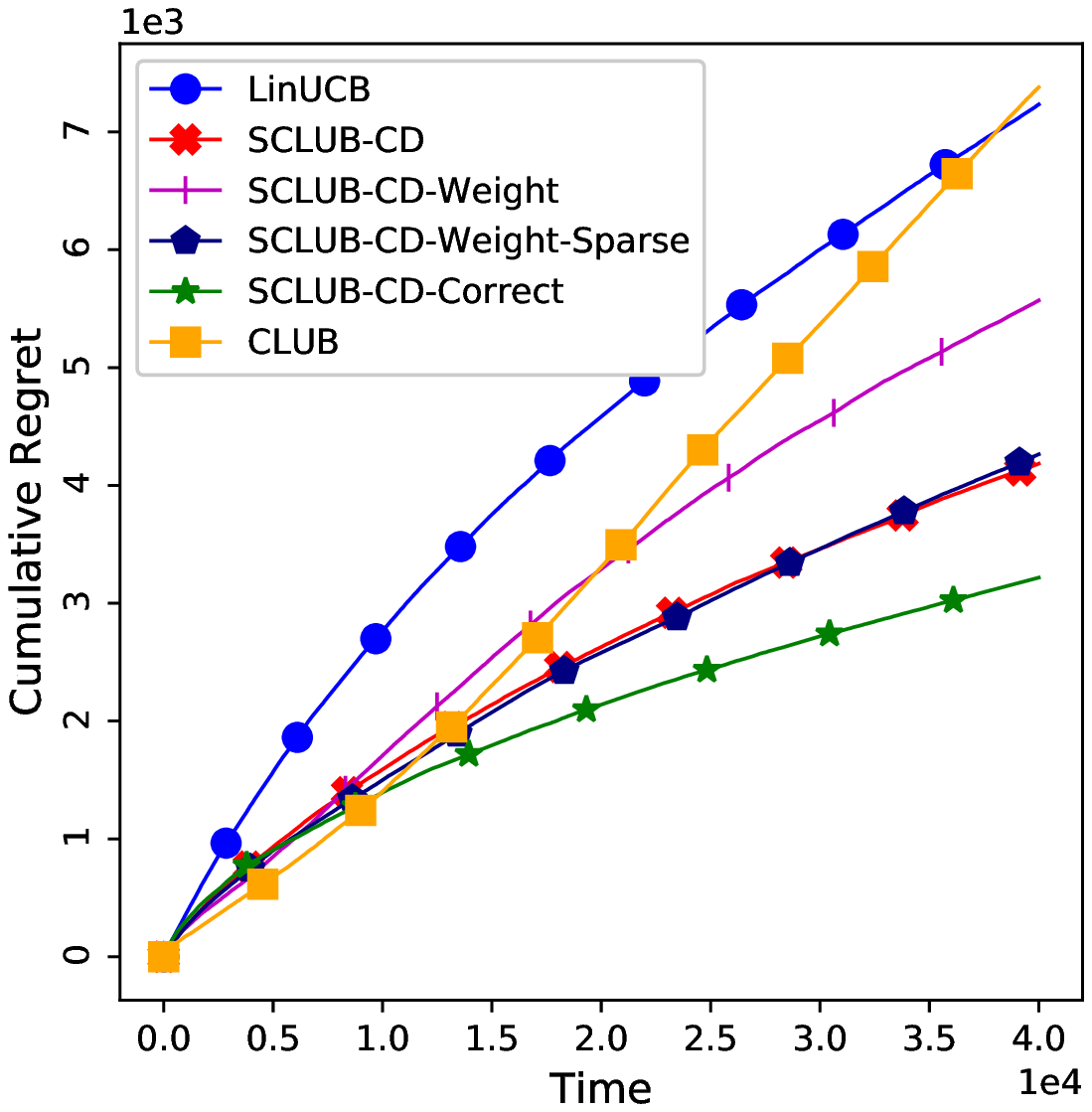}
\label{fig:data_performance_50_25}}
\hfill
\subfigure[$\sigma_{\text{c}}=0.5$, and  $\sigma_{\epsilon}=0.5$.]{ \includegraphics[width=.47\linewidth,draft=false]{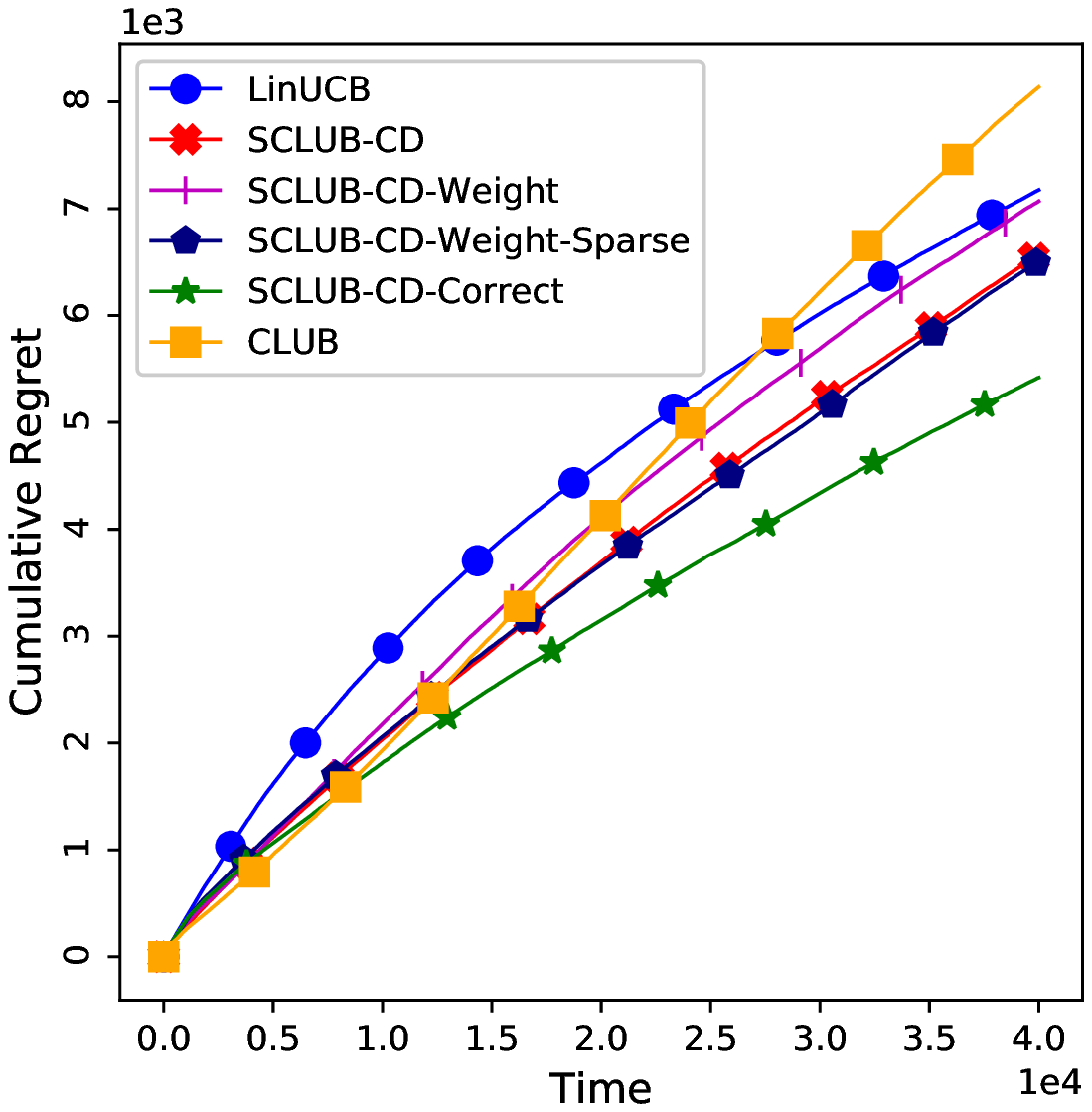}
\label{fig:data_performance_50_50}}
\caption{Cumulative regret for the synthetic dataset.
}
\label{fig:artificial_performance}
\end{figure}
%%%
\begin{figure}[t]
\centering
\subfigure{\includegraphics[width=.5\linewidth,draft=false]{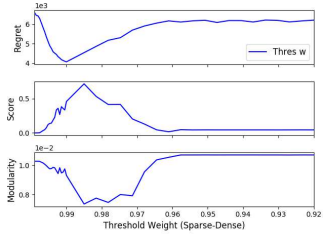}
}
\caption{Sparsity and regret}
\label{fig:sparsity_and_regret}
\end{figure}

Fig.   \ref{fig:artificial_performance} shows the cumulative regret as a function of the recommendation opportunities in the case of  synthetic dataset and under various combinations of reward noise and intra-cluster noise. In the long-term,  SCLUB-CD outperforms its competitors consistently over all scenarios, with a substantial gap (in terms of cumulative regret) for lower intra-cluster noise ($\sigma_{\text{c}}=0.25$), Fig. \ref{fig:data_performance_25_25} and Fig. \ref{fig:data_performance_50_25}. It is   interesting to observe that also when the intra-cluster noise increases  (Fig. \ref{fig:data_performance_25_50} and Fig. \ref{fig:data_performance_50_50}), clustering the user space with the proposed approach still leads to a system improvement. With respect to LinUCB, which does not cluster the users, the gain is experienced because of the faster learning process: LinUCB learns $N$ preference vector while SCLUB-CD learns $M$ ones.
With respect to CLUB, the gain is motivated by $i)$ the online learning estimation of the graph at each time opportunity,  $ii)$ the graph-based clustering not limited to identify connected components,  $iii)$ the binary and sparse modelling of the graph. 
In Fig.  \ref{fig:data_performance_50_25}, LinUCB might outperform SCLUB-CD for time horizon greater than $50000$. However,   we considered simulations  with  $N=100$, to have a fair comparison with LinUCB. By increasing the dimensionality of the user space this  potential crossing point is shifted far away in time \cite{Kaige:A18}.

Finally, the gap between SCLUB-CD and SCLUB-CD-Correct shows the potential room for improvement of the proposed algorithm, as discussed at the end of this session. The comparison with SCLUB-CD-Weight  and SCLUB-CD-Weight-Sparse shows  the gain in controlling the sparsity level (and therefore number of clusters) in the proposed algorithm. 

Fig. \ref{fig:sparsity_and_regret} shows the regret under different level of sparseness. The sparseness is controlled by the tuneable parameter $n$. The cumulative regrets are shown in the first subfigure. The second subfigure presents the clustering quality measured by NMI (Normalised Mutual Information). The the subfigure presents the corresponding modularity. It is clear that the pattern of regret mimics that of NMI. The reason is straightforward that better clustering quality (higher NMI) leads to better performance (lower regret). This result indicates the importance of constructing the user graph with an appropriate level of sparseness which leads to high clustering quality and therefore improves performance.  
\begin{figure} 
\centering
\subfigure[LastFM]{ \includegraphics[width=.47\linewidth,draft=false]{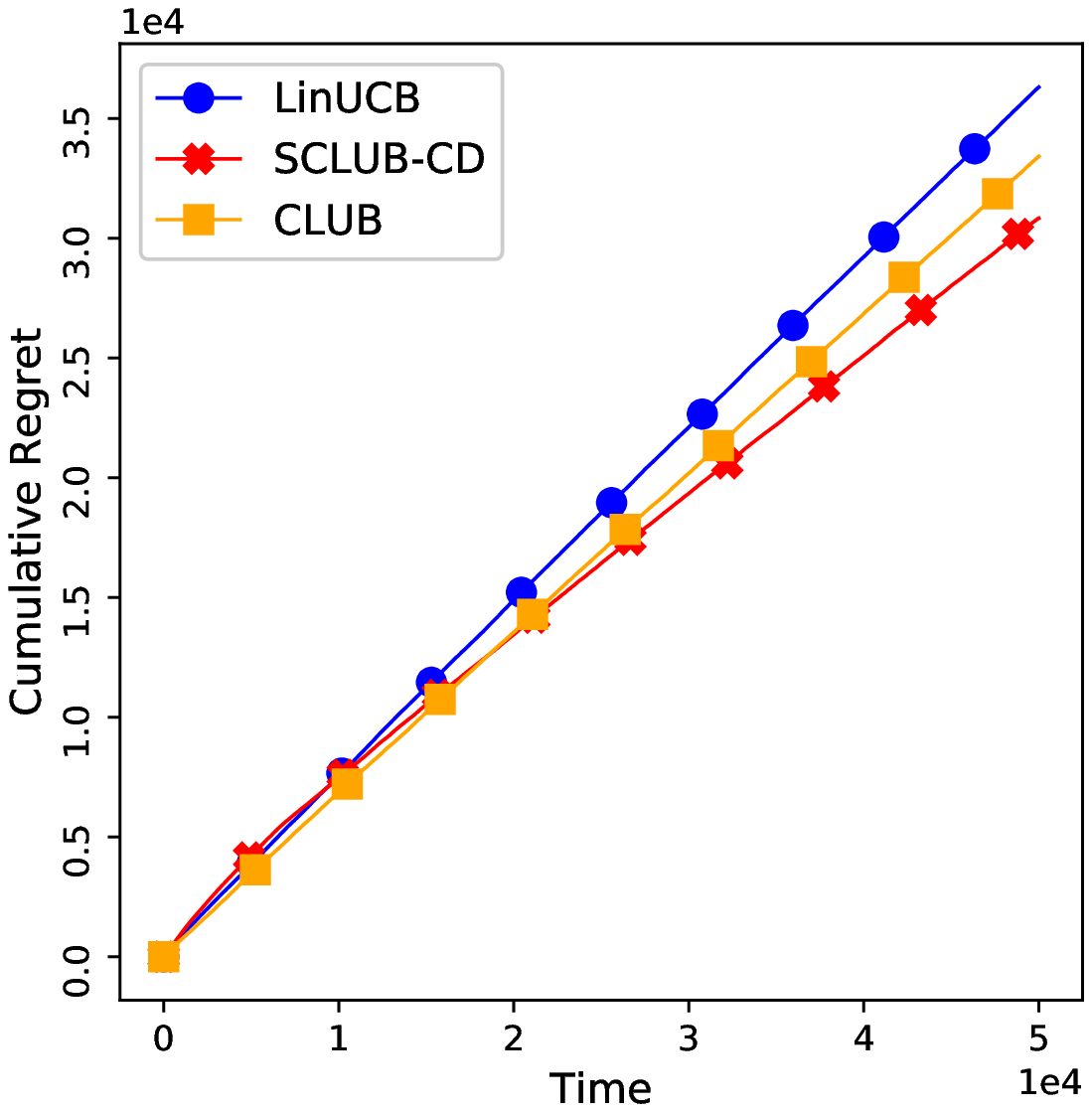}
\label{fig:lastfm_data_performance}}
\hfill
\subfigure[Delicious]{ \includegraphics[width=.47\linewidth,draft=false]{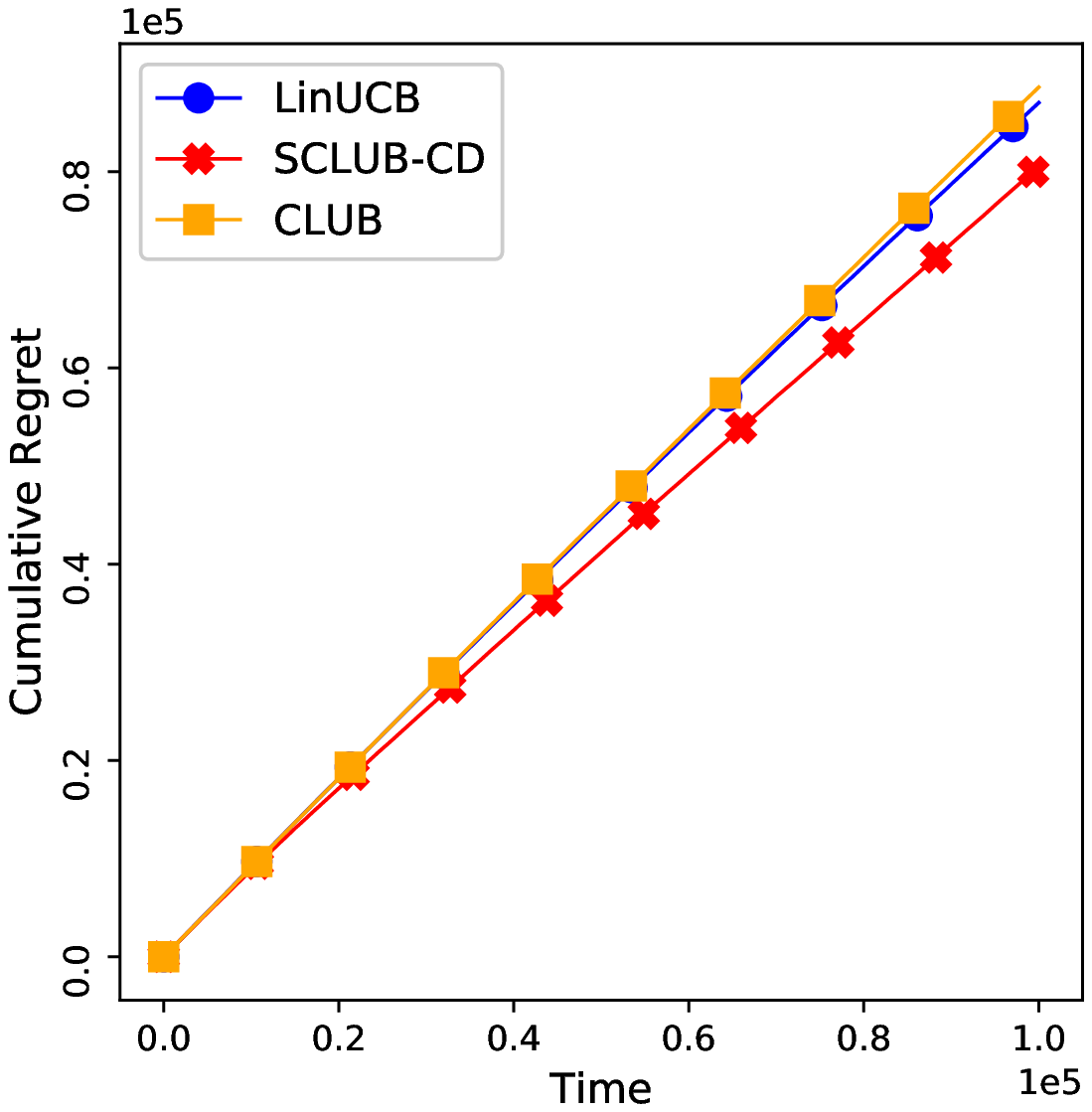}
\label{fig:delicious_data_performance}}
\subfigure[LastFM Clustering]{ \includegraphics[width=.47\linewidth,draft=false]{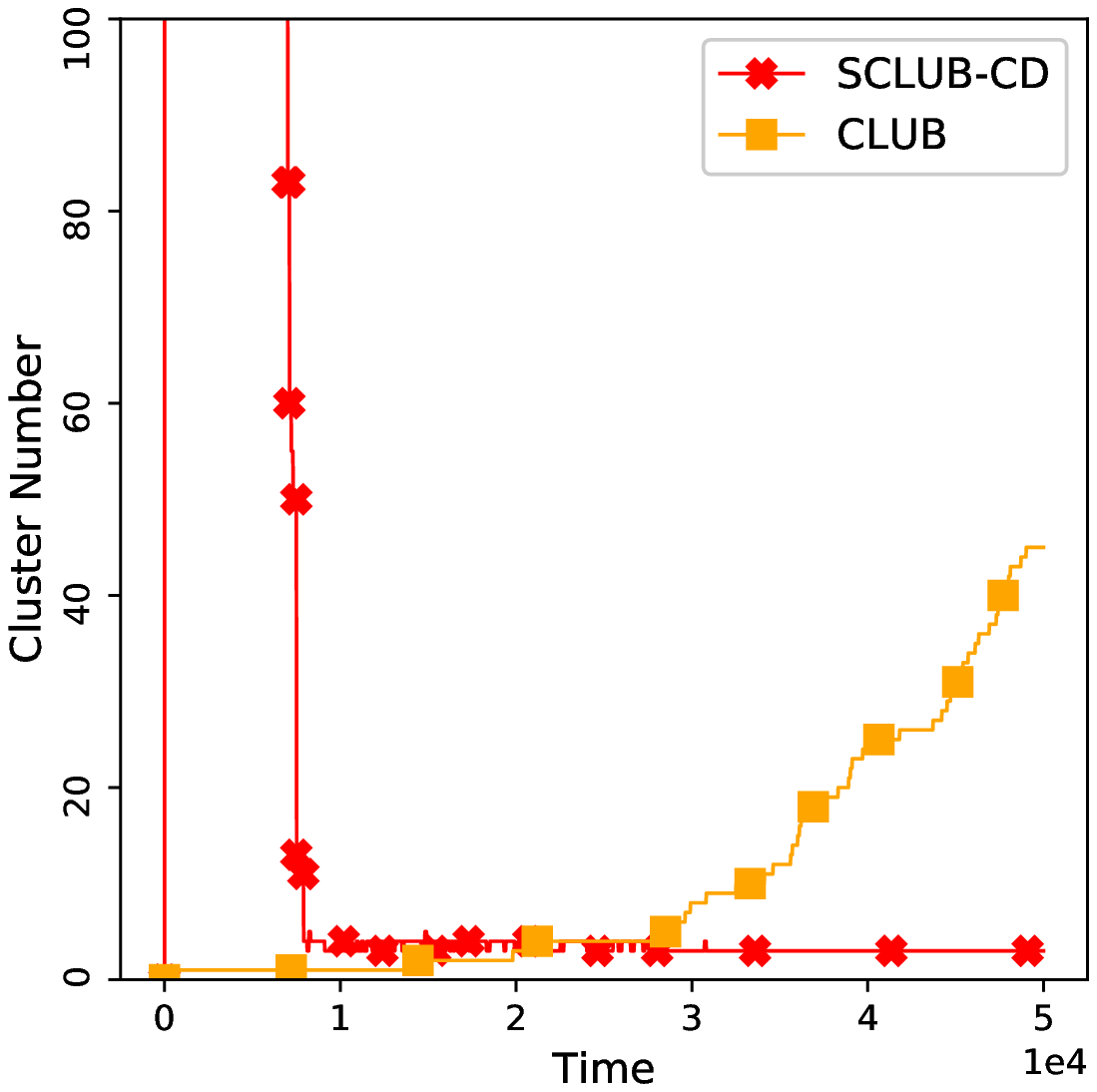}
\label{fig:lastfm_evolution_of_cluster_number}}
\hfill
\subfigure[ Delicious Clustering]{ \includegraphics[width=.47\linewidth,draft=false]{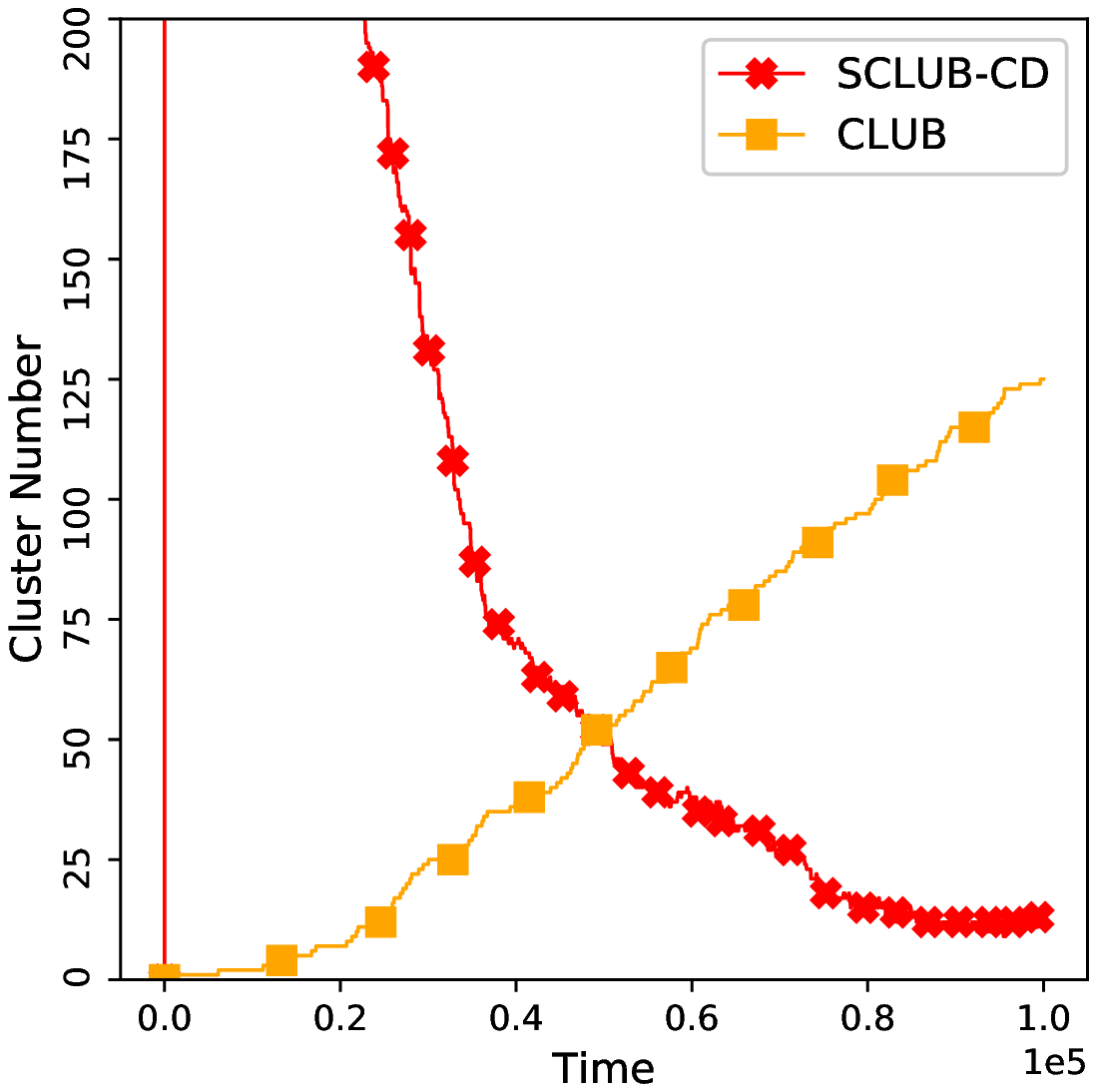}
\label{fig:delicious_evolution_of_cluster_number}}
\caption{Cumulative regret for the real-world dataset.
}
\label{fig:real_data_performance}
\end{figure}

Fig. \ref{fig:real_data_performance} presents
results on real world datasets. In \textbf{LastFM}, SCLUB-CD maintains a leading margin, this is due to a better clustering methodology. In particular,  CLUB tends to indentify many clusters, while SCLUB-CD identifies   $M=3$ user clusters only with $n=300$, as shown in Fig.\ref{fig:lastfm_evolution_of_cluster_number}. This means that the proposed approach is able to find the right tradeoff between dimensionality reduction and approximation in clustering users. 

In \textbf{Delicious}, SCLUB-CD  still outperforms baseline algorithms, but the leading margin is smaller. Results show it groups users into $M=11$ clusters with $n=700$. Overall, delicious represents the ``many-hits" scenario, in which each user is interested  in a small amount and quite dissimilar websites. This means that each user will select  few website only, therefore the agent can gather a small amount of feedbacks per user, which translates in a limited training set per user. Clustering users together allows to increase the dimension of the training set. Therefore SCLUB-CD outperforms LinUCB and CLUB, however due to the highly heterogenous scenario the approximation introduced by the clustering is affecting more the overall system. This justifies the reduced gain.   

%%%%%%%%%%%%%%%%%%%%%%%%%%%%%%%%
\section{Conclusion}
\label{conclusions}
We proposed a graph-based   bandit algorithm, which encodes users' similarity in preference by an undirected   and unweighted graph and groups users into clusters. The key aspects of the proposed algorithm are that $i)$  it adopts graph-based clustering to extract meaningful clusters, $ii)$ the unweigted graph makes the system more robust to  weights estimation errors, $iii)$ the proposed method  keeps the number of clusters limited (through $n$), unlike CLUB that has a number of clusters constantly increasing over time. All these components lead to an overall gain in terms of cumulative regret with respect to state-of-the-art algorithms. These results also opened  new  questions such as ``What is the sensitivity of the MAB algorithm to the cluster size?", ``Could we adopt graph signal processing to further improve the graph knowledge (exploiting also the smoothness of the reward function on the user graph)?" As future works, we will be addressing these open questions. 
\label{sec:conclusion}

\bibliographystyle{IEEEtran}
\bibliography{reference}

\end{document}